\documentclass[review]{elsarticle}
\usepackage[utf8]{inputenc}
\usepackage{algorithmic}
\usepackage[margin=2cm]{geometry}
\usepackage{amsmath}
\usepackage{amssymb}
\usepackage{amsfonts}
\usepackage{graphicx}
\usepackage{hyperref}
\usepackage{color,soul}
\usepackage{caption}
\usepackage{subcaption}
\usepackage{tabularx}
\usepackage{multirow}
\usepackage{hhline}
\usepackage{mathrsfs}
\begin{document}

\begin{frontmatter}
\title{ Composite Deep Network with Feature Weighting for Improved Delineation of COVID Infection in Lung CT}
\author{Pallabi Dutta\corref{cor1}}
\author{Sushmita Mitra}
\address{Machine Intelligence Unit, Indian Statistical Institute, Kolkata 700108, India.\\ Email: duttapallabi2907@gmail.com, sushmita@isical.ac.in}
\cortext[cor1]{Corresponding author}
\begin{abstract}
An early effective screening and grading of COVID-19 has become imperative towards optimizing the limited available resources of the medical facilities. An automated segmentation of the infected volumes in lung CT  is expected to significantly aid in the diagnosis and care of patients. However, an accurate demarcation of lesions remains problematic due to their irregular structure and location(s) within the lung. 
 
A novel deep learning architecture, Composite Deep network with Feature Weighting {\it (CDNetFW)}, is proposed for efficient delineation of infected regions from lung CT images. Initially a coarser-segmentation is performed directly at shallower levels, thereby facilitating discovery of robust and discriminatory characteristics in the hidden layers. The novel feature weighting module helps prioritise relevant feature maps to be probed, along with those regions containing crucial information within these maps. This is followed by estimating the severity of the disease. The deep network {\it CDNetFW} has been shown to outperform several state-of-the-art architectures in the COVID-19 lesion segmentation task, as measured by experimental results on CT slices from publicly available datasets, especially when it comes to defining structures involving complex geometries.
 \end{abstract}

\end{frontmatter}
\section{Introduction}

A critical step in the fight against COVID-19 is an effective screening of the level of infection in patients; such that those seriously affected can receive immediate treatment and care, as well as be  isolated to mitigate the spread of the virus. The gold standard screening method currently used for detecting COVID-19 cases is the reverse transcription polymerase chain reaction (RT-PCR)  testing, which is a very time-consuming, laborious, and complicated manual process. The test is uncomfortable, invasive, uses nasopharyngeal swabs, and has high false negative rates; with outcome being dependent on sampling errors and low viral load. Given that CT is  reliable to check changes in the lungs, its importance in the context of COVID-19 becomes all the more evident.

Conspicuous ground-glass opacity (GGO) and multiple mottling lesions, in the peripheral and posterior lungs on CT images, are hallmark characteristics of COVID-19 pneumonia \cite{gozes2020rapid}. While the GGO are hazy darkened spots in the lung, diffuse enough such that they do not block underlying blood vessels or lung structures, the consolidations correspond to areas of increased lung density  \cite{gozes2020rapid}. It is observed that with time these infection characteristics became more frequent, and are likely to spread across both lungs. 
Morozov {\it et al.} \cite{morozov2020chest} suggested a partially quantitative severity grading system on a scale of 0 to 4, with a step value of 25\%, based on the percentage of the diseased lung tissue. While  CT-0 represents healthy cohorts, grade CT-4 refers to those patients with affected lung regions $>$ 75\%. This scale was assigned by experts based on visual inspection of the lung CT scans of infected patients.

In order to speed up the discovery of disease mechanisms, machine learning and deep learning \cite{lecun15} can be effectively employed to detect abnormalities and extract textural features of the altered lung parenchyma; to be subsequently related to specific signatures of the COVID-19 virus. Automated segmentation of the lung region and lesion from CT can help outline the volume of interest (VOI) for a fast detection and grading of the severity of infection. 
A consistent and reproducible method, for the rapid evaluation of high volumes of screening or diagnostic thoracic CT studies, is possible with artificial intelligence.  Demarcation of affected lung tissues in the CT slices demands high precision. Deep learning enables circumventing the visual approximation by radiologists, to produce accurate decisions; particularly in the scenario of high volumes of disease cases.   

Typically deep learning requires a machine to learn automatically from raw data to discover representations needed for detection or classification. In the context of medical images, it directly uses pixel values of the images  at the input; thereby, overcoming the manual errors caused by inaccurate segmentation and/or subsequent hand-crafted feature extraction. Convolution neural networks (CNNs) \cite{lecun15} constitute one of the popular models of deep learning. Some of the commonly used deep learning models in medical applications include CNN,  ResNet \cite{he2016deep}, Res2Net \cite{gao2019res2net}, DenseNet \cite{huang2017densely}, and SegNet \cite{badrinarayanan2017segnet}. 

Custom architectures, like the $U$-Net \cite{ronneberger2015u}, have been designed for segmentation with promising results. Here the skip connections between the symmetric down-sampling (encoder) and up-sampling (decoder) paths, in the fully convolutional framework, provide local (high resolution) information in the global context during up-sampling; thereby, resulting in improved localization of the target region with more precise segmentation results. However, it is observed that the target lesion regions (like GGO) in lung CT scans of COVID-19 patients are of lower contrast with blurred edges, particularly in the early stages. They also appear in varying shapes and sizes within the lung. 
Attention modules were incorporated in the vanilla $U$-Net framework  \cite{oktay2018attention}, in order to enhance its segmentation performance. The encoder feature maps were recalibrated spatially with the aid of decoder feature maps. The attention module thus helped the network to focus only on the relevant  regions of interest in the target, while suppressing the contribution from irrelevant areas. 

 Both ResNet and Res2Net incorporate residual connections, to help combat the vanishing gradient issue, while developing deeper networks. Res2Net additionally employs parallel cascading convolution operations within a block. This helps capture multi-scale features in the input volume. The DenseNet makes use of residual connections by bridging all convolution layers with each other. It additionally promotes re-use of feature maps, with further improvement in performance. 
The $U$-Net++ \cite{zhou2018unet++} and Residual $U$-Net \cite{khanna2020deep} are also used in medical image segmentation. The $U$-Net++ inserts multiple convolution blocks between the encoding and decoding pathway of its $U-$Net backbone. This helps narrow down the semantic gap between the activation maps produced by the encoders and decoders; thereby, reducing the complexity of optimization. An exhaustive survey on the role of artificial intelligence in analysing and predicting COVID-19 can be found in  \cite{tseng2020computational,suri2021narrative}.  

A novel deep network Composite Deep network with Feature Weighting ({\it CDNetFW}) is developed for efficiently demarcating the COVID-19 infection in lung CT images. It is capable of learning generalized representations of the irregularly structured infected regions. The contribution of this research is summarized below.
\begin{itemize}
    \item A mini-segmentation architecture is introduced in the encoder branch of the composite deep network. Such coarser-level segmentation provides direct supervision to intermediate encoder layers for circumventing noisy activation and vanishing gradient.
    \item The feature weighting module focuses on the relevant activation responses. The activation maps carrying most pertinent data about the target regions are initially detected. This is  followed by identification of the spatial locations-of-interest within them. Incorporation of depth-wise and dilated convolutions helps generate the relevant re-calibrating weights for map refinement. Dilated convolutions help capture multi-scale features towards improved segmentation of lesions having different sizes and textures.
    \item A combination loss is used to handle the issue of class imbalance, while simultaneously minimizing the occurrence of False Positives and False Negatives in the output.
\end{itemize}

The rest of the paper is organized as follows. Section \ref{dlseg} briefly describes state-of-the-art literature on the use of deep learning for segmentation of COVID lesions from lung CT images. In Section \ref{meth} we introduce the proposed {\it CDNetFW}. Experimental results are discussed in Section \ref{res}. Comparative study is provided with state-of-the-art models, including the $U-$Net \cite{ronneberger2015u} and its variants like  $U-$Net++ \cite{zhou2018unet++}, Residual $U$-Net \cite{khanna2020deep}, and Attention $U-$Net \cite{oktay2018attention}, on segmentation of the lesions. The demarcated region is used to compute the percentage of infection in the lung; on the basis of which a grading of the severity of the disease is evaluated. The superiority of the {\it CDNetFW} is demonstrated on CT slices extracted from four sets of publicly available data, {\it viz.} MOSMED \cite{morozov2020mosmeddata}, MedSeg-COV-1 \cite{MedSeg2021}, MedSeg-COV-2 \cite{MedSeg22021}, and COV-CT-Lung-Inf-Seg \cite{jun2020covid}. Finally Section \ref{concl} concludes the article. 

\section{Deep Learning for Delineating Infected CT Volumes} \label{dlseg}

A retrospective, multi-centric study \cite{li2016artificial} employed the ResNet-50, with $U$-Net for segmenting the lung region from CT, to differentiate between COVID-19 viral pneumonia, Community acquired pneumonia (CAP) and other non-pneumonia images.  Similarly, a 3D DenseNet was trained in a diverse multinational cohort  \cite{harmon2020artificial} to localize parietal pleura/ lung parenchyma, followed by classification of COVID-19  pneumonia; with high sensitivity and specificity on an independent test set. Visualization of activation regions was performed using Grad-CAM to assess association of peripheral regions of the lung across variable amounts of disease burden. A novel Joint Classification and Segmentation system was designed \cite{wu2021jcs} to perform real-time and explainable diagnosis of COVID-19 cases. Annotation was provided with fine-grained pixel-level labels, lesion counts, infected areas and locations, benefiting various diagnosis aspects. With the explainable classification outputs coming from Res2Net, and the corresponding fine-grained lesion segmentation resulting from VGG-16, the system helped simplify and accelerate the diagnostic process for radiologists. 

An attention-based deep 3D multiple instance learning (AD3D-MIL) was developed \cite{han2020accurate} for the effective screening of COVID-19. A patient-level label was assigned to each 3D chest CT, which was viewed as a bag of instances. An attention-based pooling approach to 3D instances provided insight into the contribution of each instance towards the bag label. 

 The segmentation performance of the $U$-Net and SegNet was compared \cite{saood2021covid} on one out of the two datasets in MedSeg-29 \cite{MedSeg22021}, obtained from the Italian Society of Medical Interventional Radiology. While binary segmentation identified the lung tissues infected by COVID-19 virus, the multi-class segmentation delineated the different lesion pathologies. However, both exhibited poor performance in multi-class segmentation, particularly for identifying the pleural effusions, possibly due to scarcity of corresponding CT slices.

A multi-task architecture was introduced \cite{goncharov2021ct} to simultaneously perform classification and segmentation. The lung CT scans were grouped into COVID-19 infected and non-infected categories, while also ranking the infected scans in decreasing order of severity. The infected regions were first segmented to calculate the ratio of the affected lung region w.r.t. the total lung region, over each slice. The maximum value was used to compute the severity stage, as per the guidelines of the  Fleischner Society. 

The DUDA-Net \cite{xie2021duda}  employed a cascaded pair of $U$-Nets, with the first one extracting the lung region from the CT slice to make it easier for subsequent focus on the COVID-19 lesions. Attention modules with dilated convolutions helped the second $U$-Net to effectively capture multi-scale information, and re-weight relevant channels of the input feature map volume for better focus on the region of interest. This strategy helped decrease any segmentation error caused by smaller-sized lesions. However this entailed an increase in computational complexity in terms of the parameters involved.

A pair of attention modules were introduced \cite{zhao2021d2a} in the $U$-Net framework, to enhance feature map volumes along the skip connections as well as the up-sampled feature maps of the decoder. This dual attention strategy helped re-weight the feature maps, both spatially and channel-wise, for improved segmentation of the infected lung tissues in the slices. The dilated convolutions helped generate larger receptive fields.

The nCoVSegNet \cite{liu2021covid} employed two-stage transfer learning to deal with data scarcity in large annotated COVID-19 datasets. It acquired knowledge both from the ImageNet \cite{deng2009imagenet} and a lung nodule detection dataset  (LIDC-IDRI), to delineate COVID-19 lesions from lung CT scans in the MOSMED dataset. A global context-aware module, employing convolution blocks of varying sizes, helped capture multi-scale features. A Dual-Attention fusion module, incorporating both channel and spatial attention, enabled improved segmentation performance. 

The Inf-Net \cite{fan2020inf} involved a parallel partial decoder to aggregate  high-level features for generating a global segmentation map. A reverse attention mechanism, with an additional edge attention module, was incorporated for better delineation of the blurred edges of the lesions. The semi-supervised framework, requiring only a few labeled images and leveraging primarily unlabeled data, helped improve the learning ability with a higher performance. 
 
A lightweight CNN model LCOV-Net \cite{zhao2021lcovnet} employed a separable convolution operation, in lieu of the conventional convolution block, to capture features from 3D lung CT volumes; thereby, significantly reducing the model parameters to make it computationally lighter with faster training time. Attention mechanism was incorporated to recalibrate the feature maps, in order to emphasize the relevant features w.r.t. the region of interest.

\section{Composite Deep Network with Feature Weighting (CDNetFW)} \label{meth}

This section describes the {\it CDNetFW} along with its primary components, {\it viz.} the mini-segmentation network and the feature weighting mechanism. The loss function employed is also outlined, along with the performance evaluation metrics used.
\subsection{Architecture}

\begin{figure}[h]
    \centering
    \includegraphics[height = 14cm, width=9cm]{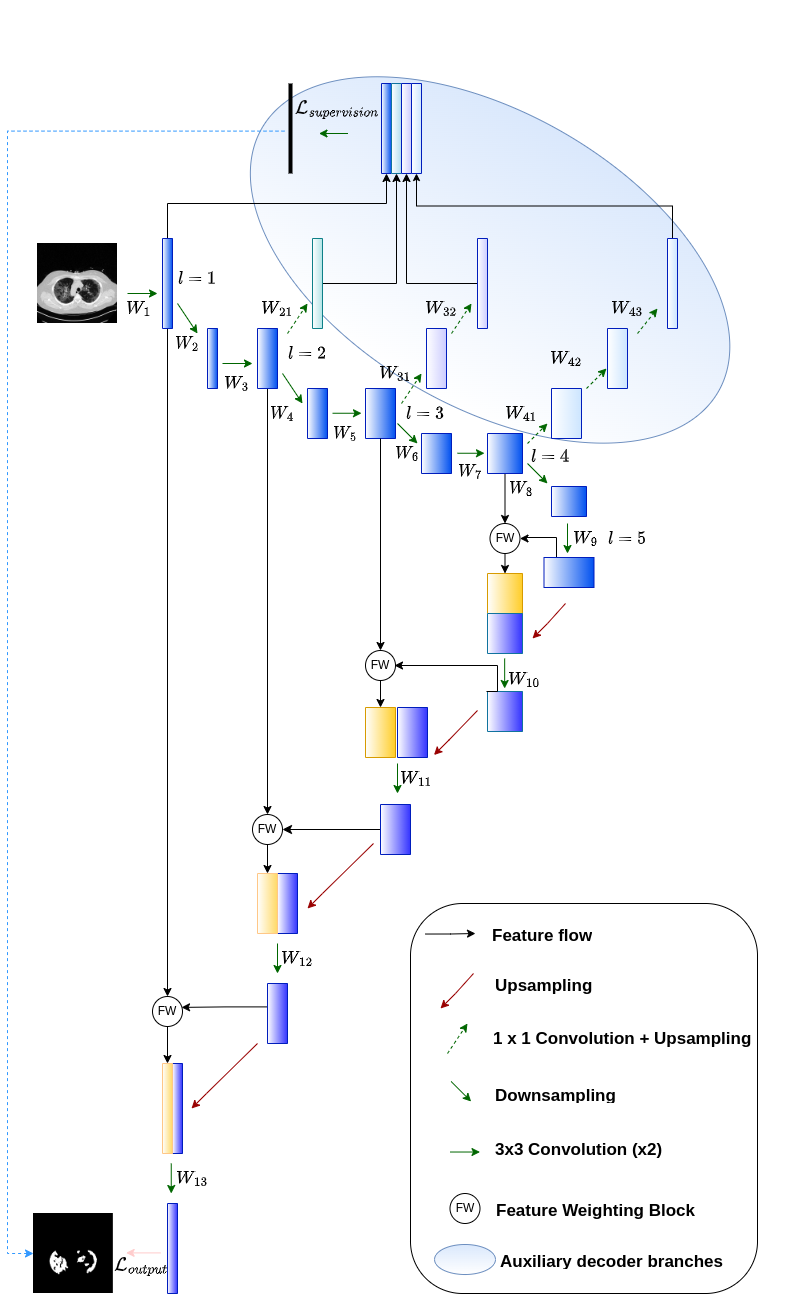}
    \caption{Schematic diagram of {\it CDNetFW}}
    \label{fig:novel unet}
\end{figure} 

\begin{figure}[h]
    \centering
    \includegraphics[height = 5cm, width=12cm]{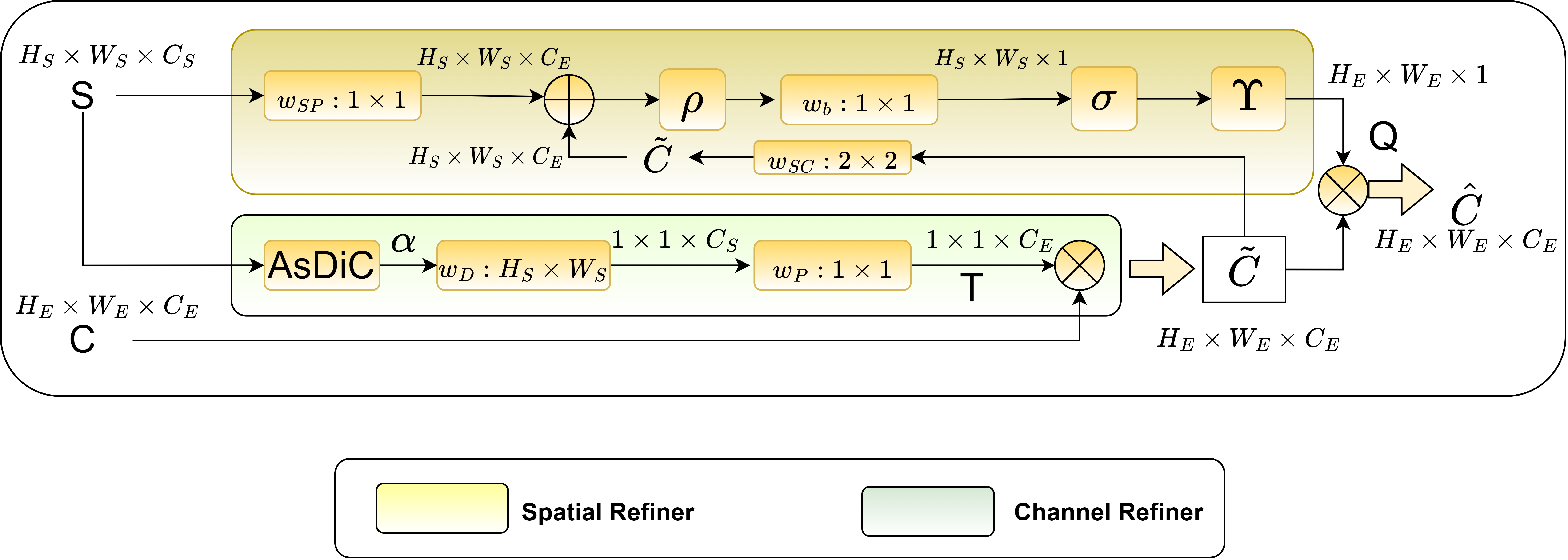}
    \caption{Feature Weighting module}
    \label{fig:feat_refine}
\end{figure}

\begin{table}[h]
\centering
\resizebox{\textwidth}{!}{
\begin{tabular}{c|l|l|l|l|c|llcccc}
\cline{1-7}
\multirow{2}{*}{\textbf{Level}} & \multicolumn{2}{c|}{\textbf{Encoder}} & \multicolumn{2}{c|}{\textbf{Decoder}}     & \multicolumn{2}{c}{Deep Supervision}                               \\\cline{2-7}
                                & Filters           & Output           & Filters             & Output             & Filters                              & Output                      \\\cline{1-7}
\multirow{5}{*}{1}              & Input Layer       & 512x512x1        & UP2D                & 512x512x128        & \multicolumn{2}{c}{\multirow{5}{*}{N/A}}                           \\
                                & Conv2D+ZP         & 512x512x64       & Conv2D+ZP           & 512x512x64         & \multicolumn{2}{c}{}                                               \\
                                & GN                & 512x512x64       & GN                  & 512x512x64         & \multicolumn{2}{c}{}                                               \\
                                & Conv2D+ZP         & 512x512x64       & Conv2D+ZP           & 512x512x64         & \multicolumn{2}{c}{}                                               \\
                                & Conv2D            & 256x256x64       & Output Layer        & 512x512x1          & \multicolumn{2}{c}{}                                               \\ \cline{1-7}
\multirow{4}{*}{2}              & Conv2D+ZP         & 256x256x128      & UP2D                & 256x256x256        & \multirow{4}{*}{Conv2D+GN+UP2D}      & \multirow{4}{*}{512x512x64} \\
                                & GN                & 256x256x128      & Conv2D+ZP           & 256x256x128        &                                      &                             \\
                                & Conv2D+ZP         & 256x256x128      & GN                  & 256x256x128        &                                      &                             \\
                                & Conv2D            & 128x128x128      & Conv2D+ZP           & 256x256x128        &                                      &                             \\ \cline{1-7}
\multirow{4}{*}{3}              & Conv2D+ZP         & 128x128x256      & UP2D                & 128x128x512        & \multirow{4}{*}{(Conv2D+GN+UP2D) x2} & \multirow{4}{*}{512x512x64} \\
                                & GN                & 128x128x256      & Conv2D+ZP           & 128x128x256        &                                      &                             \\
                                & Conv2D+ZP         & 128x128x256      & GN                  & 128x128x256        &                                      &                             \\
                                & Conv2D            & 64x64x256        & Conv2D+ZP           & 128x128x256        &                                      &                             \\ \cline{1-7}
\multirow{4}{*}{4}              & Conv2D+ZP         & 64x64x512        & UP2D                & 64x64x1024         & \multirow{4}{*}{(Conv2D+GN+UP2D) x3} & \multirow{4}{*}{512x512x64} \\
                                & GN                & 64x64x512        & Conv2D+ZP           & 64x64x512          &                                      &                             \\
                                & Conv2D+ZP         & 64x64x512        & GN                  & 64x64x512          &                                      &                             \\
                                & Conv2D            & 32x32x512        & Conv2D+ZP           & 64x64x512          &                                      &                             \\ \cline{1-7}
\multirow{3}{*}{5}              & Conv2D+ZP         & 32x32x1024       & \multicolumn{2}{c|}{\multirow{3}{*}{N/A}} & \multicolumn{2}{c}{\multirow{3}{*}{N/A}}                           \\
                                & GN                & 32x32x1024       & \multicolumn{2}{c|}{}                     & \multicolumn{2}{c}{}                                               \\
                                & Conv2D+ZP         & 32x32x1024       & \multicolumn{2}{c|}{}                     & \multicolumn{2}{c}{}     &  &  &  \\ \cline{1-7}     
\end{tabular}}
\caption{Architectural details of {\it CDNetFW}. Here Conv2D indicates 2D convolution, ZP represents zero padding, GN denotes Group Normalization, and UP2D corresponds to 2D upsampling with bilinear interpolation.}
\label{arch}
\end{table}

The architecture of {\it CDNetFW} is schematically depicted in Fig. \ref{fig:novel unet}. The CT slices are provided as input to a five-tier composite network. Unlike the conventional symmetric encoder-decoder framework used for segmentation, here the encoding path encompasses a mini-segmentation module with auxiliary decoder branches. Additional feature weighting modules are introduced to detect the  relevant activation maps from the entire set, while identifying the spatial locations-of-interest within them. This helps reduce the computational burden. The detailed framework is summarized in Table \ref{arch}.

\subsection{Mini-segmentation module}

Image segmentation in deep networks typically employs a symmetric structure of encoding and decoding pathways \cite{ronneberger2015u, oktay2018attention, khanna2020deep}. While increasing the depth of the model enhances its prediction performance, the issues like noisy feature maps at shallower levels, vanishing gradients, overfitting, etc. remain. Although skip connections are included  to tackle the problem of vanishing gradients, often less discriminative features remain prevalent (as observed from Fig. \ref{fig:sample_feature_maps_for_deep_sup}(c)). This leads to poorer generalisation capacity. In order to circumvent this problem, we propose to introduce supervision at  shallower levels of the encoder through the embedding of a mini-segmentation network within it.
Auxiliary decoder branches are attached to the second, third, and fourth levels of the encoder arm, as illustrated in Fig. \ref{fig:novel unet}. These auxiliary decoding paths provide additional supervision during training. Minimization of the loss function $\mathcal{L}_{supervision}$ at the output of this mini-segmentation network aids in enhancing the quality of the output activation at each level of the encoder. The impact of vanishing gradient is reduced by superposing the gradients obtained from these auxiliary decoder branches with that at the final output layer.

Let $\Theta = [\theta_1, \theta_2]$ consist of $\theta_1 = [W_{21},  W_{31}, W_{32}, W_{41}, W_{42}, W_{43}]$, representing the weights for the auxiliary deep supervision branches, and $\theta_2 = [W_{1}, \ldots, W_{7}]$, corresponding to the weights of the encoding path (Fig. \ref{fig:novel unet}). The adaptation is expressed as 
\begin{equation}
    \Theta = \Theta - \eta\frac{\partial \mathcal{L}_{supervision}}{\partial \Theta},
\end{equation}
where $\mathcal{L}_{supervision}$ is the auxiliary deep supervision loss and $\eta$ is the learning rate.

\subsection{Feature weighting module}

Concatenating encoder activation maps with decoder feature representations, via skip connections, helps incorporate lower-level spatial information towards improved target localization. The {\it Feature Weighting (FW)} mechanism allows focus on the relevant activation regions. Simultaneous channel and spatial refinement, incorporated in {\it FW} block, enables better segmentation of the target lesions. While channel refinement highlights feature maps of interest in the input volume, spatial refinement re-calibrates the output to highlight important locations within each such activated map. The {\it CDNetFW} employs the spatial and channel refinement mechanisms in a sequential manner, instead of applying them independently on the encoder volume to eventually combine the results as in \cite{zhao2021d2a}. Such sequential processing helps minimize redundant processing of irrelevant activation maps, while searching for spatial locations of interest within them. As a result, performance improvement occurs along with optimization of computational resources. 

\subsubsection{Channel refinement}

 \begin{figure}[h]
    \centering
    \includegraphics[height=7cm,width=6cm]{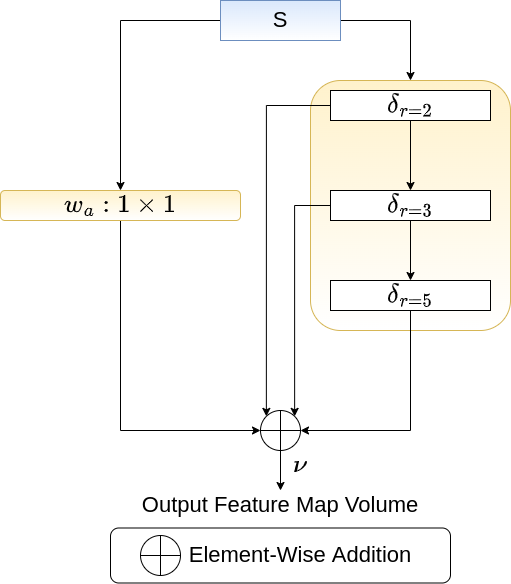}
    \caption{Assembled Dilated Convolution ({\it AsDiC}) kernels, involving different dilation rates}
    \label{fig:fig5}
\end{figure}

Consider the lower branch of the {\it FW} block at the top of Fig. \ref{fig:feat_refine}. Let $\mathbf{C} = [C^1, C^2, \ldots, C^{C_E} | C^i \in \mathbb{R}^{H_E\times W_E}]$ be the input volume from the encoder arm, where $C_E$ indicates the number of encoder channels.  Let $C_S$ correspond to the number of decoder channels, $\mathbf{W_S} = [w^1_S, w^2_S, \ldots, w_S^{C_S}]$ be the dilated convolution kernels, and $\mathbf{W}_a = [w_a^1, w_a^2, \ldots, w_a^{C_S} | w_a^i \in \mathbb{R}^{1 \times 1}]$ refer to the point-wise convolution kernels. The set of semantically rich activation maps  $\mathbf{S} = [S^1, S^2, \ldots, S^{C_S} |S^i \in \mathbb{R}^{H_S \times W_S}]$ is passed through the {\it Assembled Dilated Convolution (AsDiC)} block, involving an assembly of convolutions with varying dilation rates $r$ = 2, 3, 5 (as in Fig. \ref{fig:fig5}). The $i$th dilated convolution $\delta^i_r$ for kernel $w^i_S$ and feature map $f^j$, is expressed as
\begin{equation}
    \delta^i_r = \sum_{j=1}^{C_S}\sum_{p=-\frac{(m-1)}{2}}^{p=\frac{(m+1)}{2}}\sum_{q=-\frac{(n-1)}{2}}^{q=\frac{(n+1)}{2}}w^i_S(p,q)f^j(a+rp,b+rq).
\end{equation}
The application of multiple dilated convolutions on $\mathbf{S}$, in a pipeline of the {\it AsDiC} block, yields the output feature map volume $\pmb{\nu}$ with component $\nu^i \in \mathbb{R}^{H_S\times W_S}$ indicating the $i$th activation map. Hence 
\begin{multline}
    \nu^i = \delta_{r=2}^i(\mathbf{S}; \Omega_{\delta_{r=2}}) \oplus  \delta_{r=3}^i\{\delta_{r=2}^i (\mathbf{S}; \Omega_{\delta_{r=2}}); \Omega_{\delta_{r=3}}\} \oplus
    \delta_{r=5}^i[\delta_{r=3}^i\{\delta_{r=2}^i (\mathbf{S};\Omega_{\delta_{r=2}}); \Omega_{\delta_{r=3}}\}; \Omega_{\delta_{r=5}}] \oplus \sum_{j=1}^{C_S}w_a^k\ast S^j.  
\end{multline}
Here $\Omega_{\delta_{r=k}}$ represents the set of parameters (at dilation rate $k$) defining the dilated convolutions $\mathbf{W_S}$, with $\oplus$ indicating element-wise addition, $w_a^k$ the $k$th point-wise convolution filter and $S^j$ being the $j$th activation map in $\mathbf{S}$. Use of multiple dilated convolutions, with varying dilation rates, produces multi-scalar views from the semantically rich activation map volume at deeper levels of {\it CDNetFW}. This helps capture the appropriate characteristics of the target COVID-19 lesions, which inherently exhibit inconsistent shapes and sizes. Use of point-wise convolution, along with multiple stacked dilated convolutions, help retain information from the regions overlooked by dilated convolutions. Larger field of view, without greater convolution kernel parameters, reduces the computational cost. 

Let us now focus on the lower branch of the {\it FW} block in Fig. \ref{fig:feat_refine}. Depth-wise convolutions are applied to the output volume $\pmb{\nu} = [\nu^1, \nu^2, \ldots, \nu^{C_S} | \nu^i \in \mathbb{R}^{H_S \times W_S}]$ of {\it AsDiC} to squeeze its dimensions to $1\times 1\times C_S$, followed by a point-wise convolution for further dimensionality reduction in order to match the number of channels $C_E$ of the encoder feature maps $\mathbf{C}$. It introduces non-linearity to help learn more complex patterns. This is formulated as
\begin{equation}
    \Delta^i = w_D^i \ast \nu^i,
\end{equation}
where $\mathbf{W_D} = [w_D^1, w_D^2, \ldots, w_D^{C_E} | w_D^i \in \mathbb{R}^{H_S \times W_S}]$ represent the depth-wise convolution kernels, and $\Delta^i \in \mathbb{R}^{1 \times 1}$ is the resultant $i$th output map. The derived weights $T^i \in [0,1]$ are computed as 
\begin{equation}
    T^i = \sum_{j=1}^{C_S}w_P^k \ast \Delta^j,
\end{equation}
where $\mathbf{W}_P = [w_P^1,  w_P^2, \ldots, w_P^{C_S} | w_P^i \in \mathbb{R}^{1 \times 1}]$ are the point-wise convolution kernel filters. The output $\tilde{C} = [\tilde{C}^1, \tilde{C}^2, \ldots,  \tilde{C}^{C_E} | \tilde{C}^i \in \mathbb{R}^{H_E \times W_E}]$ is expressed as
\begin{equation}
    \tilde{C}^i = T^i \otimes C^i,
\end{equation}
where $\otimes$ refers to the element-wise multiplication involving $C^i$ (individual feature maps) of the encoder volume $\mathbf{C}$. Unlike \cite{zhao2021d2a}, which employed Global Average Pooling along with an artificial neural network to generate weights, this mechanism uses depth-wise convolutions. It aids independent learning of spatial patterns from each of the activation maps. This is important because each map is a representative of a distinct set of patterns. Besides, with such a fully convolutional module there exists no dependency on the dimensions of the input tensor; thus  making the module flexible for incorporation in other architectures. 

\subsubsection{Spatial refinement}

The $i$th spatial-refinement weight $Q^i$, illustrated in the upper branch of {\it FW} block in Fig. \ref{fig:feat_refine}, is computed as
\begin{equation}
    \mathbf{Q}^i = \Upsilon[\sigma(\sum_{l=1}^{C_E}w^k_b \ast \zeta^l)],
\end{equation}
where
\begin{equation}
    \zeta^l = \rho\{(\sum_{j=1}^{C_E}w_{SP}^k\ast S^j) \oplus (\sum_{j=1}^{C_E}w_{SC}^k\ast \tilde{C}^j)\},
\end{equation}
with $\mathbf{W_{SP}} = [w_{SP}^1,w_{SP}^2, \ldots,  w_{SP}^{C_E} | w_{SP}^i \in \mathbb{R}^{1\times 1}]$, $\mathbf{W_{SC}} =  [w_{SC}^1,  w_{SC}^2, \ldots, w_{SC}^{C_E} | w_{SC}^i \in \mathbb{R}^{2\times 2}]$ and\\ $\mathbf{W_b} =[w_b^1, w_b^2, \ldots,  w_b^{C_E} |w_b^i \in \mathbb{R}^{1\times 1}]$ being the different convolution filters characterizing the spatial refinement mechanism. Here $\rho(x): \max(0,x)$ and $\sigma(x): \frac{1}{1+e^{-x}}$ denote the ReLU and sigmoid activation functions, respectively, and $\Upsilon$ represents the up-sampling operation using bilinear interpolation.

The spatially re-calibrated output features $\mathbf{\hat{C}} = [\hat{C}^1, \hat{C}^2, \ldots, \hat{C}^{C_E} | \hat{C} \in \mathbb{R}^{H_E \times  W_E}]$ are obtained through an element-wise multiplication $\otimes$ of $\tilde{C}_{k} \in  \mathbb{R}^{C_E}$, with the weight $Q^i \in [0,1]$ derived from the decoder signal volume $\mathbf{S}$. We obtain
\begin{equation}
    \hat{C}^i = Q^i \otimes \tilde{C}_{k}.
\end{equation}
The formulation re-calibrates $\mathbf{\tilde{C}}$ by the weights generated through the spatial refinement mechanism to produce signal $\mathbf{\hat{C}}$. This helps detect the relevant activation maps, from the entire volume, followed by focusing within them  on only those locations carrying important information about the region of interest. It is unlike highlighting the relevant locations in the input encoder signal volume $\mathbf{C}$ with the generated weights $\mathbf{Q} = [Q^1, Q^2, \ldots, Q^{C_E}]$ \cite{zhao2021d2a}. There is also reduction in the computational burden.
\subsection{Loss functions}

A significant problem encountered in medical image segmentation arises due to the severe class imbalance, often existing in a target region-of-interest (to be delineated) {\it vis-a-vis} the background region. This causes the final predictions of the model to be influenced by the dominant non-target (or background) regions. Dice loss  \cite{milletari2016v} ($DL$) addresses the class imbalance issue between the foreground and background pixels. It is defined as 
\begin{multline}
     DL = 1-\left(\frac{2TP+\epsilon}{2TP+FN+FP+\epsilon}\right)
     = 1 - \left(\frac{2\sum_{i=1}^N\hat{y}_iy_i + \epsilon}{2\sum_{i=1}^N\hat{y}_iy_i + \sum_{i=1}^N y_i(1-\hat{y}_i)+\sum_{i=1}^N (1-y_i)\hat{y}_i + \epsilon}\right)\\
    =1 - \left(\frac{2 \sum_{i=1}^N \hat{y}_iy_i + \epsilon}{\sum_{i=1}^N \hat{y}_i+y_i + \epsilon}\right),
    \label{DL}
\end{multline}
where $\hat{y}_i$ and $y_i$ are the predicted and ground truth value for the $i$th pixel, respectively, $N$ is the total number of pixels, $\epsilon$ is a small random constant, $TP$, $FN$, $FP$ correspond to the true positive, false negative, false positive, respectively. However, a significant disadvantage of this loss function is that it does not account for the output imbalance, encompassing $FP$ and $FN$, in the segmentation output \cite{abraham2019novel}. \textit{Precision} specifies the proportion of pixels correctly identified as belonging to a given class (here, lesion region), relative to the total number of pixels annotated as representing that actual class. The percentage of correctly identified positive pixels (here, lesion) out of all positive pixels present in the ground truth, is represented by the metric \textit{Recall}.
\begin{equation}
    Precision=\frac{TP}{TP+FP},
    \label{prec}
\end{equation}
\begin{equation}
    Recall=\frac{TP}{TP+FN}.
    \label{reca}
\end{equation}
It is observed that training only with $DL$ results in an increased {\it Precision} at the expense of diminished {\it Recall}. This is evident from the difference in the scores of \textit{Precision} and \textit{Recall} in Table \ref{tab:loss_comp}. 

The Tversky loss ($TL$) \cite{salehi2017tversky} alleviates the problem of $FP$ and $FN$ by assigning them with weight values $\alpha$ and $\beta$, respectively.  
\begin{multline}
    TL = 1-\left(\frac{TP + \epsilon}{TP + \alpha FN + \beta FP + \epsilon}\right)
    = 1 - \frac{\sum_{i=1}^N\hat{y}_i y_i + \epsilon}{\sum_{i=1}^N\hat{y}_i y_i + \alpha\sum_{i=1}^Ny_i(1-\hat{y}_i) + \beta\sum_{i=1}^N(1-y_i)\hat{y}_i + \epsilon}.
\end{multline}

An enhanced version of $TL$, called Focal Tversky loss $FTL$ \cite{abraham2019novel}, involves an additional parameter $\gamma$ which helps target the imbalanced (smaller) Regions of Interest (ROIs) by exponentially increasing the loss for pixels with low predicted probability. This is expressed as
\begin{equation}
    FTL = (TL)^{\frac{1}{\gamma}}.
\end{equation}

We used a  linear combination $\mathscr{L}$ of the loss functions $DL$ and $FTL$, for training the proposed {\it CDNetFW}; thereby, utilizing the benefits of handling class imbalance while improving upon the balance between $Precision$ and $Recall$. Here
\begin{equation}
    \mathscr{L}=\sum_l (DL_l + FTL_l), \label{loss}
\end{equation}
for $l$ classes. Note that $l \in {0, 1}$ for binary segmentation. The total loss is computed as
\begin{equation}
    \mathcal{L}_{total\;loss} = \mathcal{L}_{supervision} + \mathcal{L}_{output}, \label{comb}
\end{equation}
where
\begin{equation}
    \mathcal{L}_{supervision} = \mathcal{L}_{output} = \mathscr{L}, \label{comb2}
\end{equation}
 The values of the hyperparameters $\alpha$, $\beta$, $\gamma$, were set at 0.7, 0.3, $\frac{4}{3}$, respectively \cite{abraham2019novel}.
 
\subsection{Performance metrics}\label{metric}

The $DSC$, $Precision$, $Recall$, $SPE$, and \textit{IoU} are employed to evaluate the segmentation produced by the {\it CDNetFW}. The $DSC$ is a harmonic mean of \textit{Precision} and \textit{Recall}, and measures the similarity between the predicted mask and the ground truth for a sample CT slice. The $SPE$ measures the ability of the model to correctly identify pixels representing the background region (here, healthy lung tissues) out of all the pixels annotated as belonging to the background. Another important metric for the segmentation task is \textit{IoU}, which calculates the amount of overlap between the predicted mask and the available ground truth. All these metrics have values lying in the range [0,1]. They are  expressed as
\begin{equation}
    DSC=\frac{2TP}{2TP+FP+FN},
    \label{dsc}
\end{equation}
\begin{equation}
    SPE=\frac{TN}{TN+FP},
\end{equation}
\begin{equation}
    IoU=\frac{TP}{TP+FP+FN}.
    \label{iou}
\end{equation}
Here \textit{TN} denotes \textit{True Negative}. A classifier is represented in terms of a graphical plot between the {\it True Positive Rate (TPR)} and {\it False Positive Rate (FPR)}, at varying thresholds of classification. {\it Receiver Operating Characteristic (ROC)} curve is  used as a comparative measure to evaluate between the classifiers. The Area Under the Curve ($AUC$) of each model is also a good evaluation measure. The model performance, under various loss functions, is depicted in terms of the Area Under the {\it Precision-Recall} curve {\it(AUC-PR)}.

\section{Experimental Results} \label{res}

Here we outline the data sets used in the study, the implementation details, qualitative and quantitative experimental results, along with their comparative analysis.

\subsection{Datasets} \label{datasets}

\begin{table}[h]
    \centering
    \caption{Datasets used}
    \resizebox{\textwidth}{!}
    {
    \begin{tabular}{c|c|c|c}
    \hline
      \textbf{No.}& \textbf{Name} & \textbf{No. of annotated samples} & \textbf{No. of slices with lesions} \\
      \hline
       1 & MOSMED \cite{morozov2020mosmeddata} & 50 & 785 \\
       \hline
       2 & MedSeg-COV-1 \cite{MedSeg2021} & $>$40 & 100\\
       \hline
       3 & MedSeg-COV-2 \cite{bell2020covid} & 9 & 373\\
       \hline
       4 & COV-CT-Lung-Inf-Seg \cite{jun2020covid} & 10 & 1351\\
       \hline
    \end{tabular}
    }
    \label{tab:my_label}
\end{table}

The COVID-19 lung CT slices were obtained from four publicly available datasets, as tabulated in Table \ref{tab:my_label}. Dataset-1  \cite{morozov2020mosmeddata} is a set of lung CT scans taken from 1110 patients belonging to five categories {\it viz.} CT-0, CT-1, CT-2, CT-3, CT-4. While CT-0 corresponds to the normal cases, the rest of the data is split into four  groups according to increased severity of infection in the lung. Here CT-1 has 684 samples with affected lung percentage of 25\% or below, CT-2 has 125 samples with affected area ranging from 25\% to 50\%, CT-3 consisting of 45 samples having 50\% to 75\% of affected lung region, and CT-4 encompassing just two samples with 75\% and above affected lung portion. Only 50 scans, belonging to CT-1, were annotated with binary masks depicting regions of GGO. The affected lung area was assigned a label ``1", and the rest of the slice (unaffected portion without the lesion, as well as the background region) were assigned the label ``0". The CT volumes with annotated masks were used for our study. Dataset-2 \cite{MedSeg2021} consists of 100 axial CT slices from $>40$ COVID-positive patients. The slices were labelled by a radiologist to demarcate three different pathalogies of COVID-19 {\it viz.} GGOs, white consolidations and pleural effusion. Dataset-3 \cite{MedSeg22021} encompasses nine volumetric lung CT scans obtained from the Italian Society of Medical Interventional Radiology. However, out of a total of 829 slices, only 373 slices are provided with annotations indicating the regions with GGOs and white consolidations. Dataset-4 \cite{jun2020covid} is a collection of lung CT scans from 20 patients, with annotations done by two radiologists which were later validated by another experienced radiologist. The ground truth of these slices consists of only two labels: ``1" and ``0" to indicate the diseased tissues and other regions (comprising healthy regions of lung and background). Here we used lung CT volumes from the first ten patients for extracting the slices in our experiments. This was because the remaining 10 samples contained non-uniform number of slices, being indicative of dissimilarity in the voxel spacing.

All the slices were resized to a dimension of $512 \times 512$. The voxel intensities of all CT volumes, from the four data sources, were clipped to make them lie in the range $[1000 \textit{ HU}, 170 \textit{ HU}]$, in order to filter out unnecessary details and noise. This was followed by intensity normalisation across the resultant multi-source database. Since all the CT slices in the volume do not contain COVID lesions, we selected only those having embedded lesions for the training. To account for the fact that not all datasets included labels for every possible COVID-19 pathology, we combined these different pathologies from Datasets 2 and 3 to create a single class representing COVID-19 lesions. The annotated samples, from the four multi-source datasets, were combined into a single database. This combined dataset was next randomly divided into five parts, for five-fold cross validation. Pooling the multi-source samples into one single dataset helped the model attain better generalization to learn the varying COVID-19 lesion structures and appearance (corresponding to different severity levels). We also utilized the additional ten patient lung masks from Dataset-4 to train the model for segmenting the lung region in the input CT slice while evaluating the severity of infection. 

\subsection{Results}

The experimental setup for {\it CDNetFW} was implemented in Python 3.9 on a 12GB NVIDIA GeForce RTX 2080 Ti GPU. Optimizer Adam was used with an initial learning rate of 0.0001 for the first 35 epochs; which was subsequently reduced to 0.00001 for the remaining 65 epochs. Learning rate was reduced by a factor of 50\%, when there was no further improvement in loss value after five consecutive epochs. Early stopping was employed to prevent overfitting. 
Both qualitative and quantitative analysis was made to evaluate the segmentation performance of the proposed {\it CDNetFW}. Comparative study with related state-of-the-art deep segmentation architectures demonstrate the superiority of our model, under different ablation experiments. The severity of infection was also estimated.

\begin{table}[h]
\centering
\caption{Effect of loss functions on the performance metrics}

\begin{tabular}{c|c|c|c|c}
\hline
\textbf{Loss functions}                 & \textbf{DSC} & \textbf{IoU} & \textbf{Precision} & \textbf{Recall} \\
\hline
Focal Tversky Loss                      & 0.8146       & 0.6964       & 0.7994             & \textbf{0.8386}          \\
\hline
Dice Loss                               & 0.8217       & 0.7042       & 0.8593             & 0.7945          \\
\hline
\textbf{Focal Tversky Loss + Dice Loss} & \textbf{0.8293}       & \textbf{0.7159}       & 0.8372             & 0.8241          \\
\hline
Focal  Loss + Dice Loss                 & 0.8192       & 0.70         & \textbf{0.8706}             & 0.7799          \\
\hline
\end{tabular}
\label{tab:loss_comp}
\end{table}

\subsubsection{Ablations}

\begin{figure}[h]
    \centering
    \includegraphics[height=7cm,width=7cm]{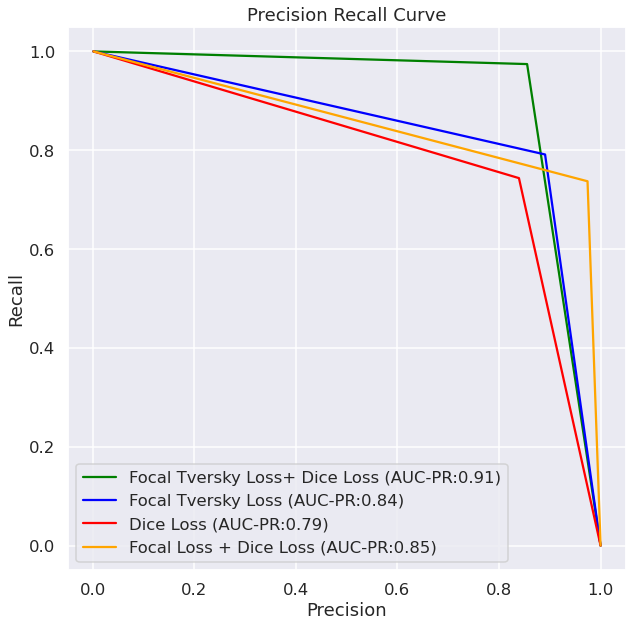}
    \caption{{\it Precision-Recall} curves for the different loss functions using the combined annotated dataset}
    \label{fig:prcurve_loss}
\end{figure}

The effect of loss functions, {\it viz.} $FTL$, $DL$, $FL + DL$, $FTL + DL$, was investigated on the performance metrics of eqns. 
(\ref{prec}), (\ref{reca}, (\ref{dsc}), (\ref{iou}), using the combined annotated dataset (as elaborated above). This is presented in Table \ref{tab:loss_comp}.
It is observed that $DSC$ and $IoU$ were the highest, when the proposed combined loss function of eqn. (\ref{loss}) was used [Row 3 in the table].  The number of pixels corresponding to infected lung region(s) is found to be significantly lower than the number of pixels from the background region, thereby resulting in major class imbalance. \\ Fig. \ref{fig:prcurve_loss} corroborates that $AUC-PR$ is the highest for the  proposed combination, signifying the correct classification of a majority of the pixels. Even in scenarios involving smaller ROIs, it can achieve a better balance between $Precision$ and $Recall$.

\begin{table}[h]
    \centering
    \caption{Effect of different modules in {\it CDNetFW} using the combined annotated dataset}
    \resizebox{\textwidth}{!}
    {
    \begin{tabular}{c|c|c|c|c|c}
        \hline
        Approach & $DSC$ & $Precision$ & $Recall$ & $SPE$ & $IoU$   \\
        \hline
        \textbf{$U-$Net backbone + auxiliary branches + {\it FW with AsDiC}} & \textbf{0.8145$\pm$0.014}  & \textbf{0.8143$\pm$0.013} & \textbf{0.8216$\pm$0.0205} & \textbf{0.9976$\pm$0.0007} & \textbf{0.6962$\pm$0.0201}\\
        \hline
        \textbf{$U-$Net backbone + auxiliary branches + {\it FW w/o AsDiC}} & 0.8006$\pm$0.019 & 0.7984$\pm$0.025 & 0.8193$\pm$0.018 & 0.9975$\pm$0.0003 & 0.6787$\pm$0.026\\
        \hline
        \textbf{$U-$Net backbone + {\it FW with AsDiC}} & 0.7698$\pm$0.033  & 0.7711$\pm$0.032 & 0.7909$\pm$0.0303 & 0.9971$\pm$0.0004 & 0.6408$\pm$0.0419\\
        \hline
        \textbf{$U-$Net backbone + {\it FW w/o AsDiC}} & 0.7356$\pm$0.023 & 0.7631$\pm$0.084 & 0.7553$\pm$0.066 & 0.997$\pm$0.0009 & 0.5983$\pm$0.03 \\
        \hline
        \textbf{$U-$Net backbone + auxiliary branches only} & 0.7839$\pm$0.009 & 0.7941$\pm$0.022 & 0.7982$\pm$0.029 & 0.9976$\pm$0.0004 & 0.6594$\pm$0.0107 \\
        \hline
        \textbf{Vanilla $U-$Net} &  0.6306 $\pm$ 0.067 & 0.7821 $\pm$ 0.072 & 0.5916 $\pm$  0.068 & 0.9969 $\pm$ 0.001 & 0.4873 $\pm$ 0.712 \\
        \hline
    \end{tabular}
    }
    \label{tab:tab_3}
\end{table}  

\begin{figure}[h]
    \centering
    \includegraphics[height = 7cm, width = 8cm]{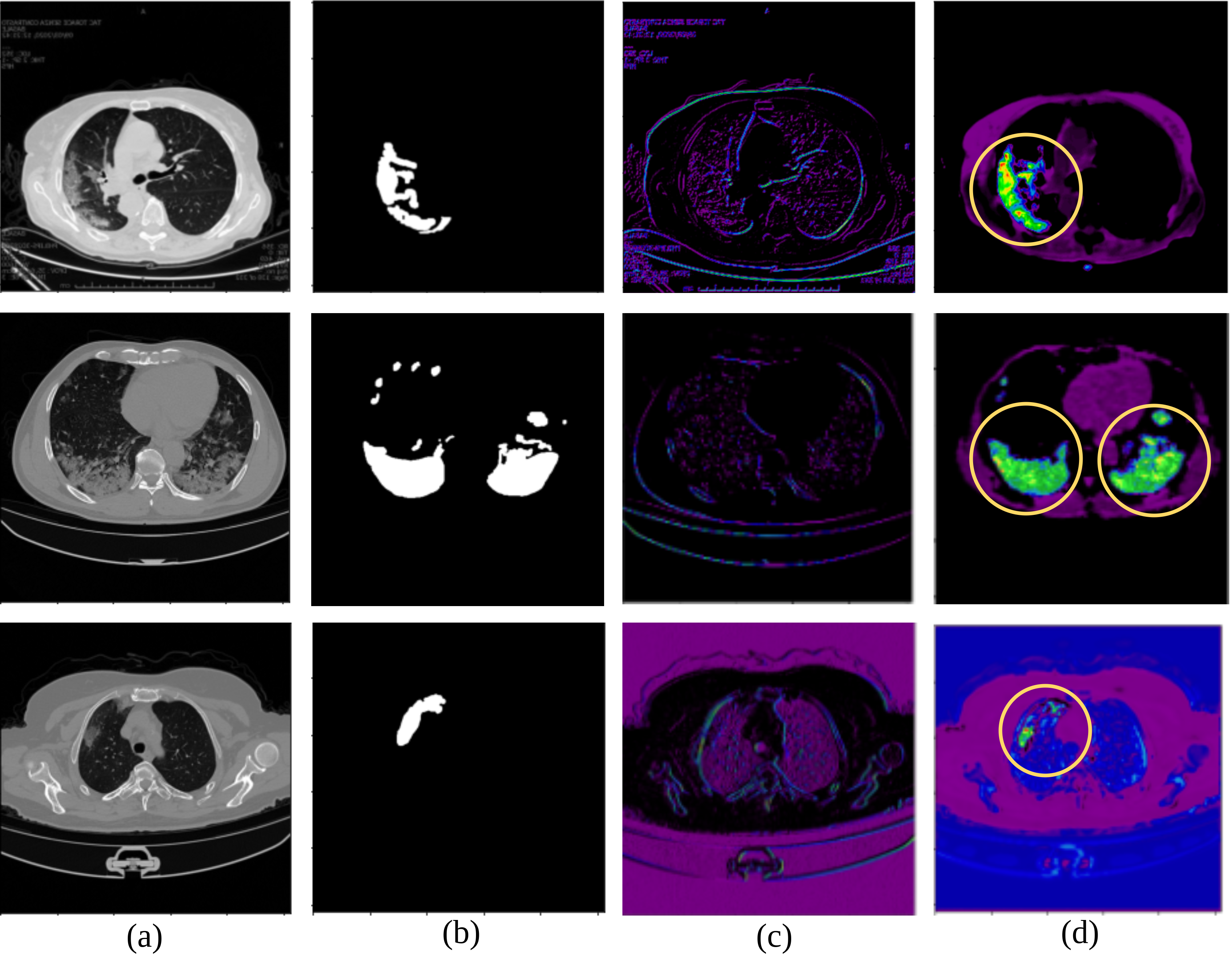}
    \caption{Impact of {\it FW} block, visualized with reference to (a) sample input CT slices, (b) corresponding ground truth, and the feature maps of {\it FW} block as displayed at its (c) input, \& (d) output.}
    \label{fig:actmap}
\end{figure}

Next some experiments were conducted to validate the role of {\it FW} and {\it AsDiC} blocks on the performance of the vanilla \textit{U-}Net using the combined annotated dataset. It is observed from Table \ref{tab:tab_3} that addition of \textit{FW} block led to a rise in \textit{DSC}. Upgrading the \textit{FW} block with the \textit{AsDiC} module resulted in a significant increase in the \textit{Recall} score; implying a decrease in $FN$ pixels. The role of feature weighting is further highlighted in Fig. \ref{fig:actmap} in terms of sample feature maps at the input and output of the {\it FW} block. It is evident that regions corresponding to the infected lung tissues get prominently displayed within the output feature maps (highlighted by yellow circles) in column (d) of the figure after passing through the {\it FW} module. 

The incorporation of \textit{auxiliary branches} to the basic \textit{U-}Net framework resulted in an increment by a margin of $>$10 \% in the \textit{DSC} and \textit{IoU}. A significant rise in both \textit{Precision} and \textit{Recall} are indicative of a simultaneous decrease in $FN$. Inclusion of {\it auxiliary branches}, in conjunction with the \textit{FW} block, further enhanced the performance of {\it CDNetFW}; with precise segmentation, as expressed by highest \textit{DSC} and \textit{Recall} scores in Row 1 of Table \ref{tab:tab_3}.

\begin{figure}[h]
    \centering
    \includegraphics[height = 7cm, width = 12cm]{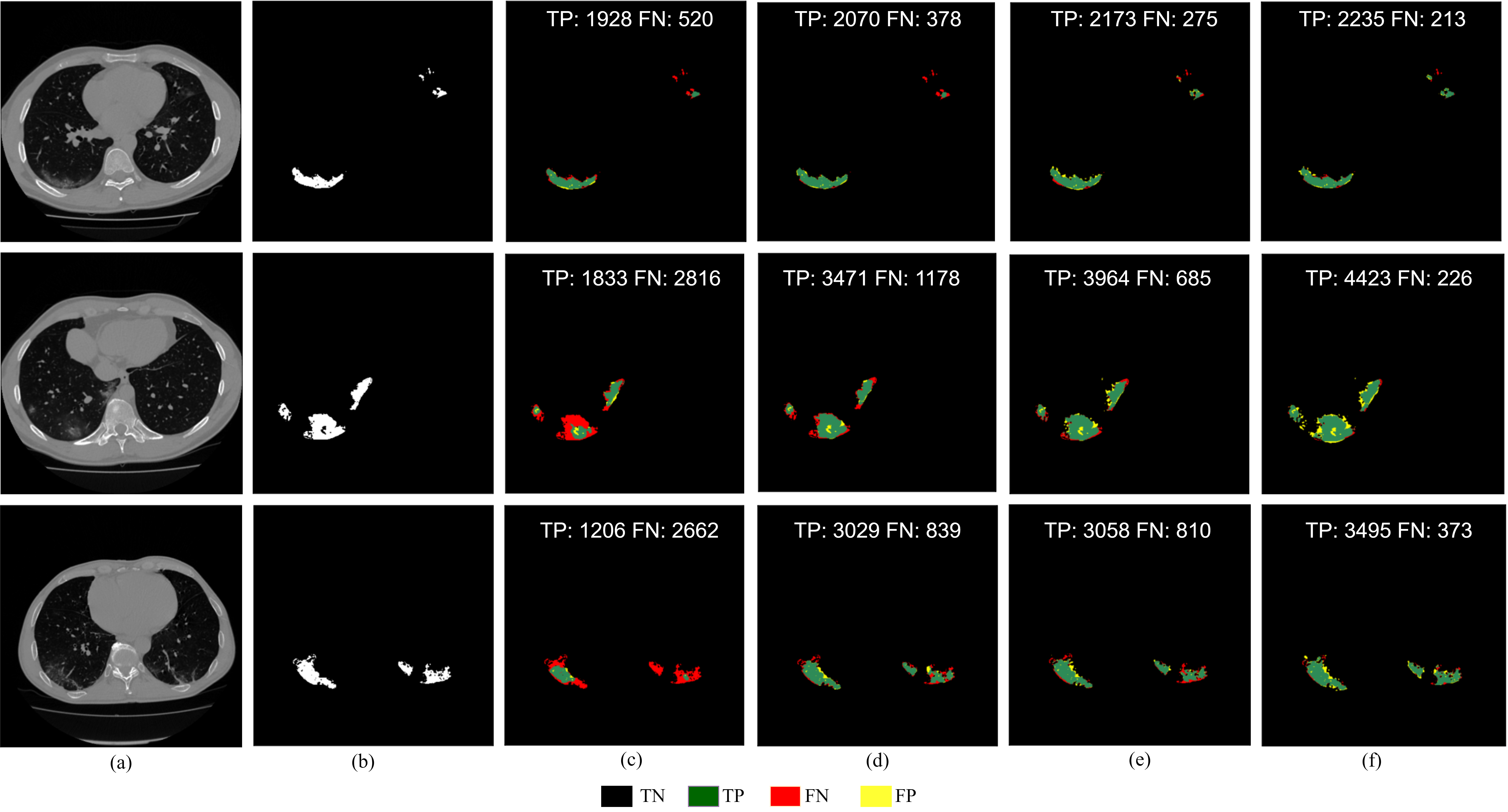}
    \caption{Segmentation on sample patients, using combined annotated dataset, depicting (a) input CT slice having (b) ground truth mask of lesions, and corresponding output generated by (c) vanilla $U$-Net, with incorporation of frequency refinement by (d) $FW$ module {\it w/o AsDiC}, (e) $FW$ module {\it with AsDiC}, and (f) auxiliary decoder branches.}
    \label{fig:abl_study}
\end{figure}

Fig. \ref{fig:abl_study} presents a visualization of the decrease in pixel count of $FN$ pixels with respect to the segmentation output generated by the vanilla \textit{U}-Net, as the different modules are added to the {\it CDNetFW}. It is observed that by incorporating both spatial and channel refinement mechanisms (in terms of $FW$ and $AsDiC$ modules), a better segmentation is obtained (both qualitatively and quantitatively) with reference to the ground truth. Adding auxiliary decoder branches generated the best outcome, as evident from part (f) of the figure.

\begin{figure}[h]
    \centering
    \includegraphics[height= 7cm, width=7cm]{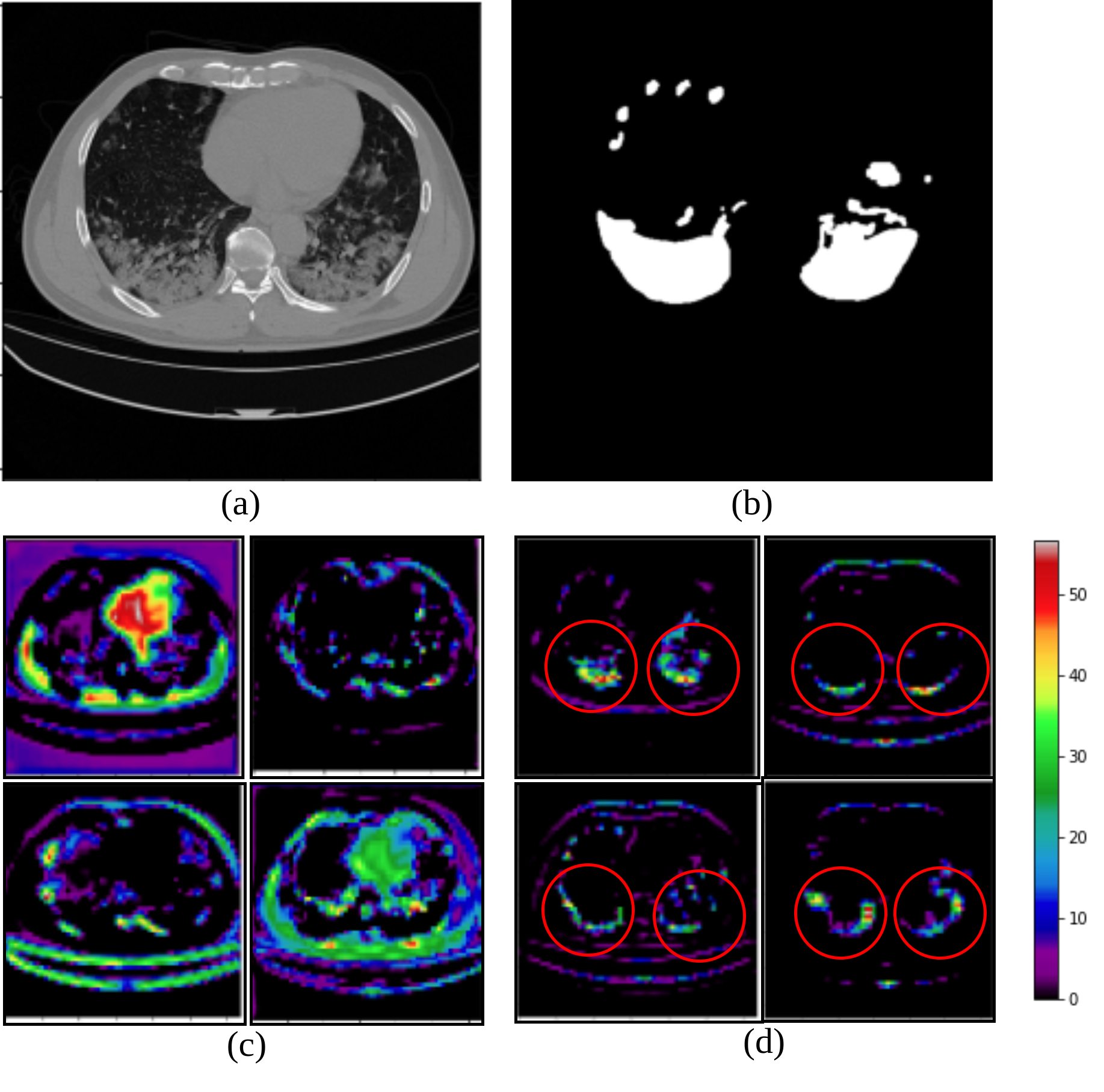}
    \caption{Feature maps from the fourth level of the encoder branch, using the combined annotated dataset. Sample (a) input CT slice, with (b) ground truth, and corresponding feature maps set (c) {\it without} auxiliary decoder branches, and (d) with auxiliary decoder branches}
    \label{fig:sample_feature_maps_for_deep_sup}
\end{figure}

Inclusion of auxiliary decoder branches (at the encoder arm) generated feature map volumes containing activation maps, which focused on relevant patterns and features in the ROI, as observed from Fig. \ref{fig:sample_feature_maps_for_deep_sup}. The marked areas within the circles of part (d) of the figure are found to represent the extracted features corresponding to the anatomy of the lesions (with reference to the ground truth).

\subsubsection{Comparisons}

\begin{table}[h]
\centering
\caption{Comparative study, over performance metrics, with related state-of-the-art architecture}
\resizebox{\textwidth}{!}{
    \begin{tabular}{c|c|c|c|c|cc}
        \hline
        Models & $DSC$ & $Precision$ & $Recall$ & $SPE$ & $IoU$   \\
        \hline
        \textbf{{\it CDNetFW}} & \textbf{0.8145$\pm$0.014}  & \textbf{0.8143$\pm$0.013} & \textbf{0.8216$\pm$0.0205} & \textbf{0.9976$\pm$0.0007} & \textbf{0.6962$\pm$0.0201}\\
        \hline
        \textbf{Attention $U$-Net \cite{oktay2018attention}} & 0.6587 $\pm$ 0.043  & 0.7537 $\pm$ 0.0751 & 0.6445 $\pm$ 0.038 & 0.9946 $\pm$ 0.006 & 0.52156 $\pm$ 0.041\\
        \hline
        \textbf{$U$-Net \cite{ronneberger2015u}} &  0.6306 $\pm$ 0.067 & 0.7821 $\pm$ 0.072 & 0.5916 $\pm$  0.068 & 0.9969 $\pm$ 0.001 & 0.4873 $\pm$ 0.712 \\
        \hline
        \textbf{$U$-Net++} \cite{zhou2018unet++} & 0.6881 $\pm$ 0.039 & 0.7209 $\pm$ 0.038 & 0.7083 $\pm$ 0.048 &  0.9961 $\pm$ 0.001 & 0.5456 $\pm$ 0.039 \\
        \hline
        \textbf{Residual $U$-Net} \cite{khanna2020deep} & 0.6501 $\pm$ 0.018 & 0.7837 $\pm$ 0.047 & 0.6004 $\pm$ 0.044 &  0.9980 $\pm$ 0.0006 & 0.5123 $\pm$ 0.017 \\
        \hline
    \end{tabular}}
    \label{tab:quan_res}
\end{table}

Table \ref{tab:quan_res} presents a comparative study of {\it CDNetFW} with state-of-the-art models, like  \textit{U-Net}, \textit{U-Net++}, \textit{Attention U-Net} and \textit{Residual $U$-Net}, employing five-fold cross validation over the various performance metrics $DSC$, $Recall$, $IoU$, $Precision$ and $SPE$ (based on the combined annotated dataset). The best scores are marked in bold in the table. It is evident that {\it CDNetFW} outperforms the remaining models. 

This reiterates the contention that by
extending the vanilla \textit{U-Net} framework with the novel feature weighting modules encompassing \textit{FW} and {\it AsDiC}, along with the \textit{auxiliary decoder branches}, enables {\it CDNetFW} to accurately identify the pixels belonging to the COVID-19 lesion category with a smaller number of $FN$s. 
By focusing only on the significant features that correspond to the target lesion regions, our architecture is found to be more effective in separating the target region(s) from a CT slice. 

The values of \textit{DSC} and \textit{IoU}, the most widely used metrics of evaluation in medical image segmentation, demonstrate discernible improvement; thereby, establishing the efficacy of {\it CDNetFW}. Significant improvement in both \textit{Precision} and \textit{Recall} metrics indicate greater detection of $TP$ pixels, with simultaneous reduction in the $FP$s. This validates the effectiveness of  {\it CDNetFW} in capturing the structure of the infection lesions encompassing blurred edges with irregular contours and shapes.

\begin{figure}[h]
    \centering
    \includegraphics[height=10cm,width=13cm]{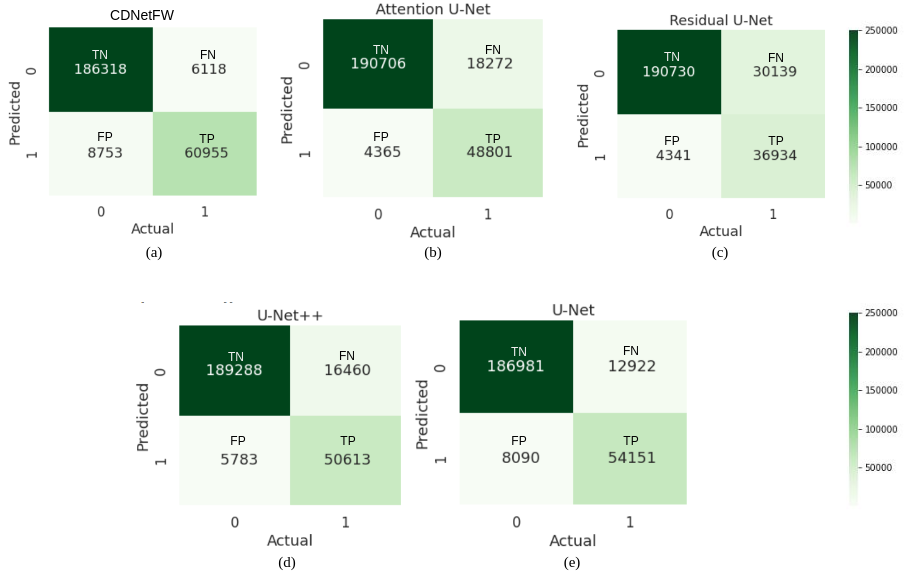}
    \caption{Confusion matrix, with combined annotated dataset, generated by (a) {\it CDNetFW}, (b) Attention $U$-Net, (c)  Residual $U$-Net, (d) $U$-Net++, and (e) vanilla $U$-Net. }
    \label{fig:confmat}
\end{figure}

\begin{figure}[h]
    \centering
    \begin{subfigure}[b]{0.45\textwidth}
         \centering
         \includegraphics[height=7.5cm, width=9cm]{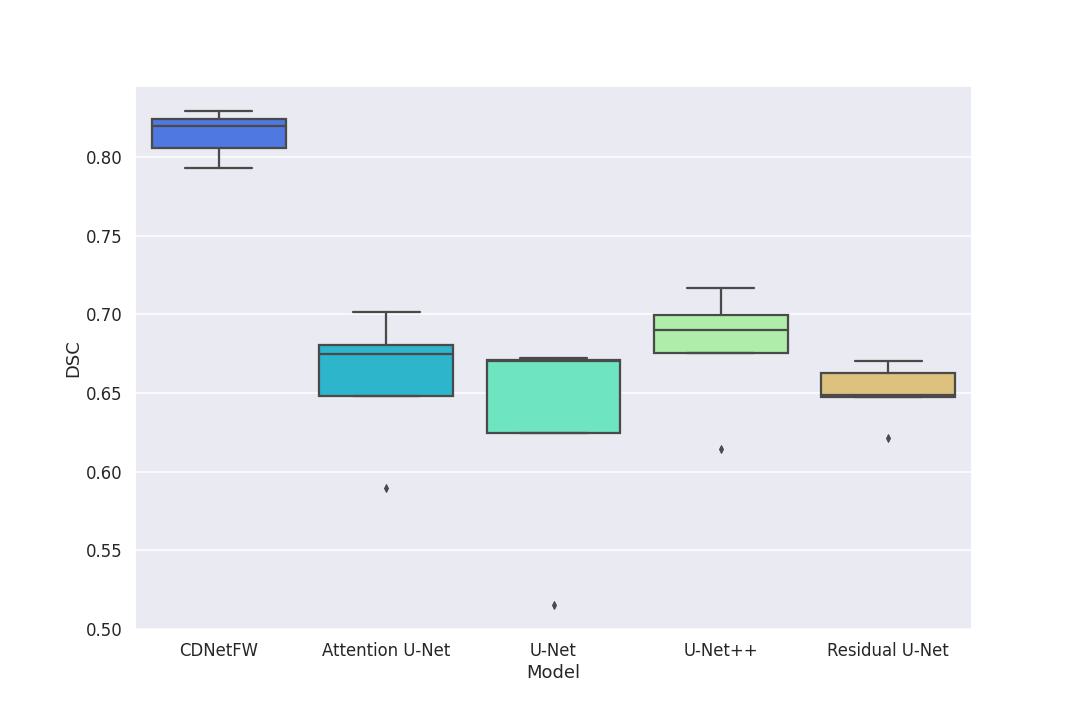}
         \caption{}
         \label{fig:boxplot}
     \end{subfigure}
     \begin{subfigure}[b]{0.45\textwidth}
         \centering
         \includegraphics[height=7cm,width=9cm]{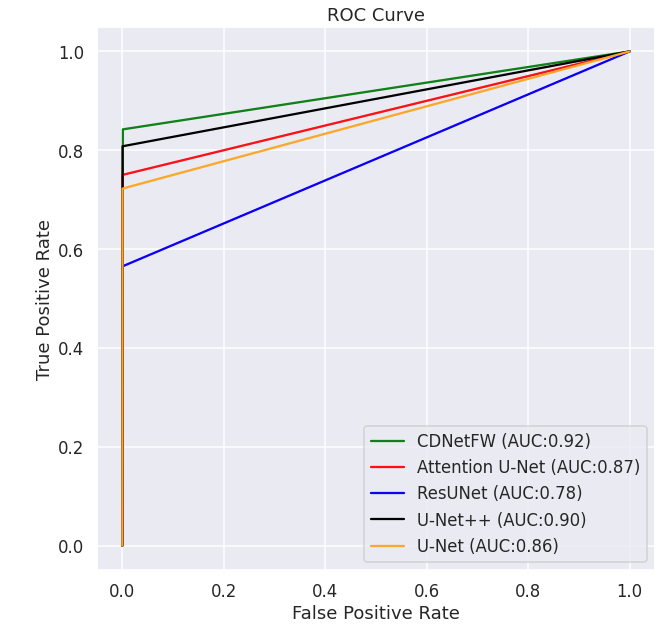}
         \caption{}
          \label{fig:rocplot}
     \end{subfigure}
    \caption{Comparative study with related models, on the combined annotated dataset, using (a) Box plot  in terms of $DSC$, and (b) $ROC$ curves with $AUC$.}
    \label{box-roc}
\end{figure}

Results are corroborated in terms of corresponding confusion matrices in Fig. \ref{fig:confmat}. The box plot of $DSC$ and the $ROC$ curve with $AUC$ in Fig. \ref{box-roc} additionally emphasize the superiority of our {\it CDNetFW}. 

\begin{figure}[h]
    \centering
    \includegraphics[height=14cm,width=14cm]{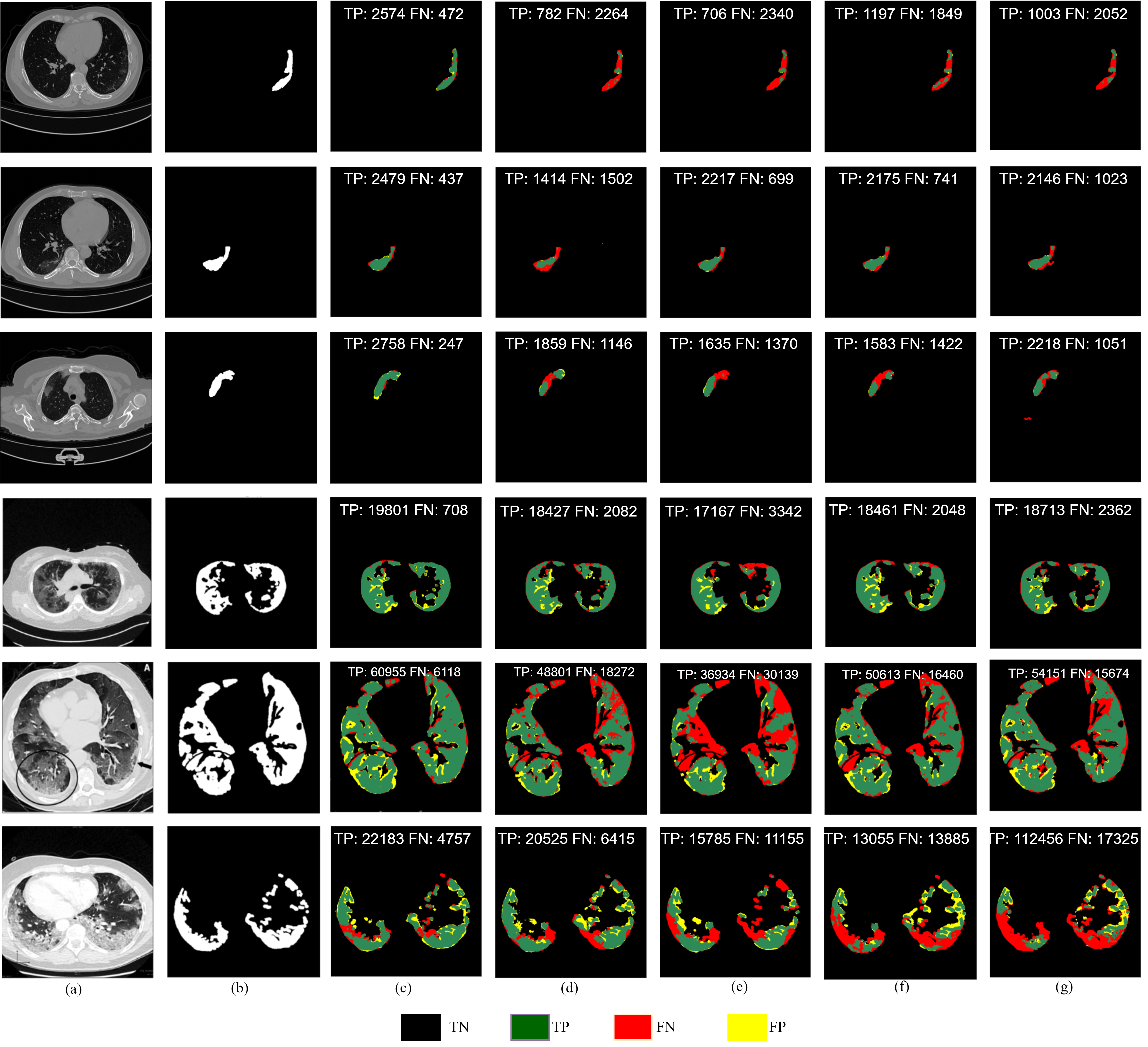}
    \caption{Qualitative comparison in segmentation output of {\it CDNetFW} with related state-of-the-art deep learning architectures, using the combined annotated dataset. (a) Sample CT slice, (b) the corresponding ground truth of the lesion, with predicted segmentation by (c) {\it CDNetFW}, (d) Attention $U-$Net (e) $U-$Net, (f) $U-$Net++, and (g) Residual $U-$Net.}
    \label{fig:qual_results}
\end{figure}

Fig. \ref{fig:qual_results} displays a qualitative study of the segmentation maps (on six sample CT scans), generated by our {\it CDNetFW}, and compared with respect to the related state-of-the-art models (referred above). It is evident that the vanilla $U$-Net, Attention $U$-Net, Residual $U$-Net, and $U$-Net++ are unable to extract the entire lesion region appropriately. Presence of $FN$ pixels can be clearly viewed in the corresponding maps. On the other hand, the segmentation map of the {\it CDNetFW} indicates the highest concentration of $TP$ pixels. It is visually evident that the {\it CDNetFW} outperforms the compared baseline architectures by more accurately segmenting the affected lung tissue, as observed from the sample CT slices.
\begin{table}[h]
\centering
    \caption{Comparative study with recent literature, in segmenting COVID lesions from lung CT}    
\resizebox{\textwidth}{!}{
\begin{tabular}{l|l|l|l|c}
\hline
\textbf{Model} & \textbf{Characteristics} & \multicolumn{2}{c|}{\textbf{Dataset}} & \textbf{DSC}   \\
\hline
    &     & \textbf{Train Set} & \textbf{Test Set}    &       \\
\hline
{\it CDNetFW}       & Composite deep network,   & 80\% of CT images from & 20\% of CT images from  & \textbf{0.82}  \\
 & with feature weighting &   MOSMED \cite{morozov2020mosmeddata},  & MOSMED \cite{morozov2020mosmeddata},  & \\
 & & MedSeg-COV-1 \cite{MedSeg2021}, & MedSeg-COV-1 \cite{MedSeg2021}, & \\
 & & MedSeg-COV-2 \cite{bell2020covid} and & MedSeg-COV-2 \cite{bell2020covid} and &\\
  & & COV-CT-Lung-Inf-Seg \cite{jun2020covid} & COV-CT-Lung-Inf-Seg \cite{jun2020covid} & \\
\hline
{\it Inf-Net}         & Parallel partial decoder to  & 45 randomly selected CT images & 50 CT images from & 0.682 \\
 & generate coarse output, & from   MedSeg-COV-1 \cite{MedSeg2021} & MedSeg-COV-1 \cite{MedSeg2021}, &\\
 & with refinement by reverse & & & \\
 & and edge attention & & & \\
\hline
Goncharov et al & Multi-task approach for & MOSMED \cite{morozov2020mosmeddata}, & MOSMED \cite{morozov2020mosmeddata} & 0.63  \\
 & segmentation with classification &  MedSeg-COV-2 \cite{bell2020covid}, & &\\
 & & COV-CT-Lung-Inf-Seg \cite{jun2020covid} & &\\
\hline
{\it D2A-Net} & Dual-attention and  & MedSeg-COV-2 \cite{bell2020covid}, & MedSeg-COV-1 \cite{MedSeg2021} & 0.72  \\
 & hybrid dilated convolutions &  COV-CT-Lung-Inf-Seg \cite{jun2020covid} & & \\
\hline
{\it nCoVSegNet}      & Transfer learning,  & 40 cases from MOSMED \cite{morozov2020mosmeddata},  & 10 cases from MOSMED \cite{morozov2020mosmeddata},  & 0.69  \\
& channel and spatial attention & LIDC-IDRI & COV-CT-Lung-Inf-Seg \cite{jun2020covid} & \\
\hline
{\it LCOV-Net}        & Lightweight CNN   & 80\% of CT images  & 20\% of CT images & 0.78 \\
& with attention & from 10 private hospitals & from 10 private hospitals & \\
\hline
\end{tabular}}
\label{tab:comp1}
\end{table}

Finally a comparative tabulation of a few other related state-of-the-art models (from literature), dealing with delineation of COVID lesions from lung CT, is provided in Table \ref{tab:comp1} in terms of $DSC$. The superiority of the proposed {\it CDNetFW} is observed. 

\subsubsection{Severity of infection}

\begin{figure}[t]
    \centering
    \includegraphics[height = 11cm, width = 8cm]{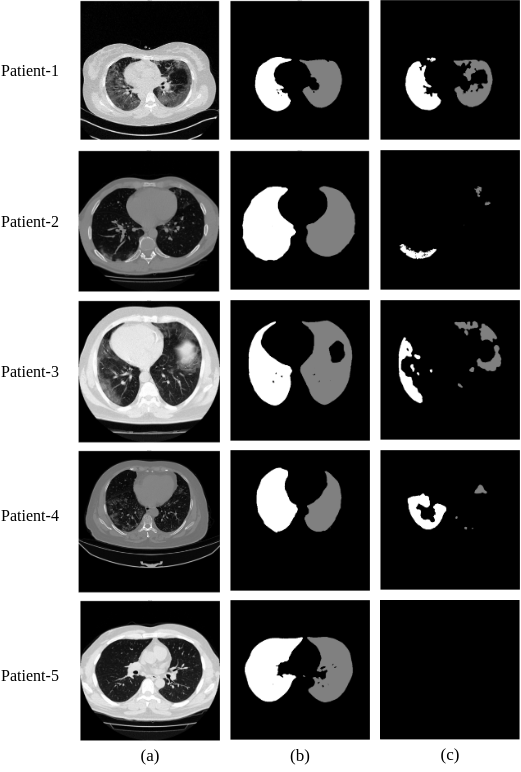}
    \caption{Gradation of severity in sample COVID-infected patients, depicting the (a) lung CT slice, (b) lung mask (white and gray indicating left and right lung, respectively), and the corresponding delineated lesion regions from both lungs.}
    \label{fig:severity}
\end{figure}

The severity grading of a sample patient was computed, based on the infection region in the corresponding lung CT. The predicted lung mask and the lesion segmentation mask were used to calculate the ratio of the affected lung area to total lung area, individually for the left and right lung. The maximum of these two ratios determined the grade of severity of infection in the patient \cite{morozov2020chest}. The severity grades are assigned according the following criteria: (i) CT-0: Healthy patients (ii) CT-1: Patients with infected lung \% $<$25 (iii) CT-2: Patients with infected lung \% between 25-50 (iv) CT-3: Patients with infected lung \% between 50 and 75 (v) CT-4: Patients with infected lung \%  $>$75
Fig. \ref{fig:severity} depicts the sample lung CT slices of five different patients, highlighting their maximum visible lung area with reference to the total CT volume in each case.  The ratio computation results in the following prediction.
\begin{itemize}
\item Patient-1: Affected Left Lung \% = 81.92, Affected Right Lung \% = 69.00, Grade: CT-4;
\item Patient-2: Affected Left Lung \% = 7.75, Affected Right Lung \% = 2.27, Grade: CT-1; \item Patient-3: Affected Left Lung \% = 31.00, Affected Right Lung \% = 23.06, Grade: CT-2; \item Patient-4: Affected Left Lung \% = 38.36, Affected Right Lung \% = 6.40, Grade: CT-2; \item Patient-5: Affected Left Lung \% = Nil, Affected Right Lung \% = Nil, Grade: CT-0.
\end{itemize}

\section{Conclusions and Discussion} \label{concl}

A novel composite deep architecture {\it CDNetFW}, which integrates a mini-segmentation network with feature weighting mechanism, has been developed for the accurate segmentation of COVID lesions from lung CT slices. The {\it FW} and {\it AsDiC} modules contribute towards feature refinement for effective delineation of irregular shapes and contours of lesions, their low contrast, and associated blurring. The network allows highlighting important activation maps, while focusing on key locations within them; thereby, allowing {\it CDNetFW} to accurately delineate the relevant lesions. Auxiliary branches enable learning of stronger and more discriminating features at shallower levels of the network. This enhances the output segmentation quality. Class imbalance and false negatives are tackled using a combination loss, while outlining the lesion structure. This also leads to accurate computation of the affected lung area to help grade the severity of infection.

In the absence of sufficient annotated samples, the model could not be trained to distinguish between the different pathologies, like GGO, white consolidation, and pleural effusion. The research is currently being extended to multi-class segmentation. It can result in an effective tool for the screening and early detection of patients who may have contracted the COVID-19 pathogen. In individual patients infected with the virus, and demonstrating associated pulmonary abnormalities, the same methodologies can be used to accurately and more rapidly assess disease progression and guide therapy with effective patient management. 
Volumetric modeling of the lung can help in determining the optimal prognostics for recovery, based on the existing damage. Incorporation of additional genomic inputs is expected to throw further illumination on the investigation. 

\section*{Acknowledgments}
\label{sec:acknowledgments}
This work was supported by the J. C. Bose National Fellowship, sanction no. JCB/2020/000033 of S. Mitra.

\bibliographystyle{IEEEtran}
\bibliography{covid}

\begin{thebibliography}{10}
\providecommand{\url}[1]{#1}
\csname url@samestyle\endcsname
\providecommand{\newblock}{\relax}
\providecommand{\bibinfo}[2]{#2}
\providecommand{\BIBentrySTDinterwordspacing}{\spaceskip=0pt\relax}
\providecommand{\BIBentryALTinterwordstretchfactor}{4}
\providecommand{\BIBentryALTinterwordspacing}{\spaceskip=\fontdimen2\font plus
\BIBentryALTinterwordstretchfactor\fontdimen3\font minus
  \fontdimen4\font\relax}
\providecommand{\BIBforeignlanguage}[2]{{%
\expandafter\ifx\csname l@#1\endcsname\relax
\typeout{** WARNING: IEEEtran.bst: No hyphenation pattern has been}%
\typeout{** loaded for the language `#1'. Using the pattern for}%
\typeout{** the default language instead.}%
\else
\language=\csname l@#1\endcsname
\fi
#2}}
\providecommand{\BIBdecl}{\relax}
\BIBdecl

\bibitem{gozes2020rapid}
\BIBentryALTinterwordspacing
O.~Gozes, M.~Frid-Adar, and {\it et al.}, ``Rapid {AI} development cycle for
  the {C}oronavirus ({COVID-19}) pandemic: {I}nitial results for automated
  detection \& patient monitoring using deep learning {CT} image analysis,''
  \emph{ar{X}iv:2003.05037}, 2020. [Online]. Available:
  \url{https://spectrum.ieee.org/hospitals-deploy-ai-tools-detect-covid19-chest-scans}
\BIBentrySTDinterwordspacing

\bibitem{morozov2020chest}
S.~P. Morozov, V.~Y. Chernina, and {\it et al.}, ``Chest computed tomography
  for outcome prediction in laboratory-confirmed {COVID-19}: A retrospective
  analysis of 38,051 cases,'' \emph{Digital Diagnostics}, vol.~1, no.~1, pp.
  27--36, 2020.

\bibitem{lecun15}
Y.~LeCun, Y.~Bengio, and {\it et al.}, ``Deep learning,'' \emph{Nature}, vol.
  521, pp. 436\--444, 2015.

\bibitem{he2016deep}
K.~He, X.~Zhang, and {\it et al.}, ``Deep residual learning for image
  recognition,'' in \emph{Proceedings of the IEEE {C}onference on {C}omputer
  {V}ision and {P}attern {R}ecognition}, 2016, pp. 770\--778.

\bibitem{gao2019res2net}
S.~H. Gao, M.~M. Cheng, and {\it et al.}, ``Res2{N}et: {A} new multi-scale
  backbone architecture,'' \emph{IEEE {T}ransactions on {P}attern {A}nalysis
  and {M}achine {I}ntelligence}, vol.~43, pp. 652\--662, 2019.

\bibitem{huang2017densely}
G.~Huang, Z.~Liu, and {\it et al.}, ``Densely connected convolutional
  networks,'' in \emph{Proceedings of the IEEE {C}onference on {C}omputer
  {V}ision and {P}attern {R}ecognition}, 2017, pp. 4700\--4708.

\bibitem{badrinarayanan2017segnet}
V.~Badrinarayanan, A.~Kendall, and {\it et al.}, ``Seg{N}et: A deep
  convolutional encoder-decoder architecture for image segmentation,''
  \emph{{IEEE} {T}ransactions on {P}attern {A}nalysis and {M}achine
  {I}ntelligence}, vol.~39, pp. 2481\--2495, 2017.

\bibitem{ronneberger2015u}
O.~Ronneberger, P.~Fischer, and {\it et al.}, ``{\it U}-{N}et: {C}onvolutional
  networks for biomedical image segmentation,'' in \emph{Proccedings of the
  {I}nternational {C}onference on {M}edical {I}mage {C}omputing and {C}omputer
  {A}ssisted {I}ntervention, ({MICCAI})}.\hskip 1em plus 0.5em minus
  0.4em\relax Springer, 2015, pp. 234\--241.

\bibitem{oktay2018attention}
O.~Oktay, J.~Schlemper, and {\it et al.}, ``Attention {$U$-N}et: {L}earning
  where to look for the pancreas,'' in \emph{Proceedings of the {M}edical
  {I}maging with {D}eep {L}earning}, 2018.

\bibitem{zhou2018unet++}
Z.~Zhou, R.~Siddiquee, and {\it et al.}, ``U{N}et++: A nested {U}-{N}et
  architecture for medical image segmentation,'' in \emph{Deep {L}earning in
  {M}edical {I}mage {A}nalysis and {M}ultimodal {L}earning for {C}linical
  {D}ecision {S}upport}.\hskip 1em plus 0.5em minus 0.4em\relax Springer, 2018,
  pp. 3\--11.

\bibitem{khanna2020deep}
A.~Khanna, N.~D. Londhe, and {\it et al.}, ``A deep {R}esidual {U}-{N}et
  convolutional neural network for automated lung segmentation in computed
  tomography images,'' \emph{Biocybernetics and {B}iomedical {E}ngineering},
  vol.~40, no.~3, pp. 1314--1327, 2020.

\bibitem{tseng2020computational}
V.~S. Tseng, J.~J.~C. Ying, and {\it et al.}, ``Computational intelligence
  techniques for combating {COVID-19}: {A} survey,'' \emph{IEEE {C}omputational
  {I}ntelligence {M}agazine}, vol.~15, pp. 10\--22, 2020.

\bibitem{suri2021narrative}
\BIBentryALTinterwordspacing
J.~S. Suri, S.~Agarwal, and {\it et al.}, ``A narrative review on
  characterization of acute respiratory distress syndrome in
  {COVID-19}-infected lungs using artificial intelligence,'' \emph{Computers in
  {B}iology and {M}edicine}, vol. 130, p. 104210, 2021. [Online]. Available:
  \url{https://doi.org/10.1016/j.compbiomed.2021.104210}
\BIBentrySTDinterwordspacing

\bibitem{morozov2020mosmeddata}
S.~P. Morozov, A.~E. Andreychenko, and {\it et al.}, ``M{OSMED} data: {D}ata
  set of 1110 chest {CT} scans performed during the {COVID-19} epidemic,''
  \emph{Digital {D}iagnostics}, vol.~1, pp. 49\--59, 2020.

\bibitem{MedSeg2021}
\BIBentryALTinterwordspacing
MedSeg, H.~B. Jenssen, and {\it et al.}, ``{MedSeg Covid Dataset 1},'' 1 2021.
  [Online]. Available:
  \url{https://figshare.com/articles/dataset/MedSeg_Covid_Dataset_1/13521488}
\BIBentrySTDinterwordspacing

\bibitem{MedSeg22021}
\BIBentryALTinterwordspacing
------, ``Med{S}eg {COVID} {D}ataset 2,'' 2021. [Online]. Available:
  \url{https://figshare.com/articles/dataset/Covid_Dataset_2/13521509}
\BIBentrySTDinterwordspacing

\bibitem{jun2020covid}
\BIBentryALTinterwordspacing
J.~Ma, C.~Ge, and {\it et al.}, ``{COVID-19} {CT} {L}ung and {I}nfection
  {S}egmentation {D}ataset,'' Apr. 2020. [Online]. Available:
  \url{https://doi.org/10.5281/zenodo.3757476}
\BIBentrySTDinterwordspacing

\bibitem{li2016artificial}
\BIBentryALTinterwordspacing
L.~Li, L.~Qin, and {\it et al.}, ``Artificial intelligence distinguishes
  {COVID-19} from community acquired pneumonia on chest {CT},''
  \emph{Radiology}, vol. 296, 2020. [Online]. Available:
  \url{https://doi.org/10.1148/radiol.2020200905}
\BIBentrySTDinterwordspacing

\bibitem{harmon2020artificial}
\BIBentryALTinterwordspacing
S.~A. Harmon, T.~H. Sanford, and {\it et al.}, ``Artificial intelligence for
  the detection of {COVID-19} pneumonia on chest {CT} using multinational
  datasets,'' \emph{Nature {C}ommunications}, vol.~11, 2020. [Online].
  Available: \url{https://doi.org/10.1038/s41467-020-17971-2}
\BIBentrySTDinterwordspacing

\bibitem{wu2021jcs}
\BIBentryALTinterwordspacing
Y.~H. Wu, S.~H. Gao, and {\it et al.}, ``{JCS}: {A}n explainable {COVID-19}
  diagnosis system by joint classification and segmentation,'' \emph{{IEEE}
  {T}ransactions on {I}mage {P}rocessing}, vol.~30, pp. 3113\--3126, 2021.
  [Online]. Available: \url{https://doi.org/10.1109/TIP.2021.3058783}
\BIBentrySTDinterwordspacing

\bibitem{han2020accurate}
Z.~Han, B.~Wei, and {\it et al.}, ``Accurate screening of {COVID-19} using
  attention-based deep 3{D} multiple instance learning,'' \emph{IEEE
  {T}ransactions on {M}edical {I}maging}, vol.~39, pp. 2584\--2594, 2020.

\bibitem{saood2021covid}
A.~Saood and I.~Hatem, ``{COVID-19} lung {CT} image segmentation using deep
  learning methods: {$U$-N}et versus {S}eg{N}et,'' \emph{{BMC} {M}edical
  {I}maging}, vol.~21, pp. 1\--10, 2021.

\bibitem{goncharov2021ct}
\BIBentryALTinterwordspacing
M.~Goncharov, M.~Pisov, and {\it et al.}, ``{CT-B}ased {COVID-19} triage:
  {D}eep multitask learning improves joint identification and severity
  quantification,'' \emph{Medical {I}mage {A}nalysis}, vol.~71, p. 102054,
  2021. [Online]. Available: \url{https://doi.org/10.1016/j.media.2021.102054}
\BIBentrySTDinterwordspacing

\bibitem{xie2021duda}
F.~Xie, Z.~Huang, and {\it et al.}, ``{DUDA-N}et: {A} double {U}-shaped dilated
  attention network for automatic infection area segmentation in {COVID-19}
  lung {CT} images,'' \emph{{I}nternational {J}ournal of {C}omputer {A}ssisted
  {R}adiology and {S}urgery}, vol.~16, pp. 1425\--1434, 2021.

\bibitem{zhao2021d2a}
\BIBentryALTinterwordspacing
X.~Zhao, P.~Zhang, and {\it et al.}, ``{D2A $U$-N}et: {A}utomatic segmentation
  of {COVID-19 CT} slices based on dual attention and hybrid dilated
  convolution,'' \emph{Computers in {B}iology and {M}edicine}, vol. 135, p.
  104526, 2021. [Online]. Available:
  \url{https://doi.org/10.1016/j.compbiomed.2021.104526}
\BIBentrySTDinterwordspacing

\bibitem{liu2021covid}
J.~Liu, B.~Dong, and {\it et al.}, ``{COVID-19} lung infection segmentation
  with a novel two-stage cross-domain transfer learning framework,''
  \emph{Medical {I}mage {A}nalysis}, vol.~74, p. 102205, 2021.

\bibitem{deng2009imagenet}
J.~Deng, W.~Dong, and {\it et al.}, ``Image{N}et: A large-scale hierarchical
  image database,'' in \emph{Proceedings of {IEEE} {C}onference on {C}omputer
  {V}ision and {P}attern {R}ecognition}, 2009, pp. 248\--255.

\bibitem{fan2020inf}
D.~P. Fan, T.~Zhou, and {\it et al.}, ``Inf-{N}et: Automatic {COVID-19} lung
  infection segmentation from {CT} images,'' \emph{{IEEE} {T}ransactions on
  {M}edical {I}maging}, vol.~39, pp. 2626\--2637, 2020.

\bibitem{zhao2021lcovnet}
Q.~Zhao, H.~Wang, and {\it et al.}, ``{LCOV-NET}: {A} lightweight neural
  network for {COVID-19} pneumonia lesion segmentation from {3D CT} images,''
  in \emph{Proceedings of the {IEEE} 18th {I}nternational {S}ymposium on
  {B}iomedical {I}maging ({ISBI})}, 2021, pp. 42\--45.

\bibitem{milletari2016v}
F.~Milletari, N.~Navab, and {\it et al.}, ``V-net: {F}ully convolutional neural
  networks for volumetric medical image segmentation,'' in \emph{Proceedings of
  {F}ourth {I}nternational {C}onference on 3D vision (3DV)}.\hskip 1em plus
  0.5em minus 0.4em\relax IEEE, 2016, pp. 565\--571.

\bibitem{abraham2019novel}
N.~Abraham and N.~M. Khan, ``A novel focal {T}versky loss function with
  improved attention {U}-{N}et for lesion segmentation,'' in \emph{Proceedings
  of {IEEE} 16th {I}nternational {S}ymposium on {B}iomedical {I}maging {(ISBI
  2019)}}, 2019, pp. 683\--687.

\bibitem{salehi2017tversky}
S.~S.~M. Salehi, D.~Erdogmus, and {\it et al.}, ``Tversky loss function for
  image segmentation using 3{D} fully convolutional deep networks,'' in
  \emph{Proceedings of {I}nternational {W}orkshop on {M}achine {L}earning in
  {M}edical {I}maging}.\hskip 1em plus 0.5em minus 0.4em\relax Springer, 2017,
  pp. 379\--387.

\bibitem{bell2020covid}
\BIBentryALTinterwordspacing
``{COVID-19} {CT} {S}egmentation {D}ataset,'' 2020. [Online]. Available:
  \url{http://medicalsegmentation.com/covid19/}
\BIBentrySTDinterwordspacing

\end{thebibliography}
\end{document}